\documentclass[prd,superscriptaddress,floatfix,amsmath,footinbib,amssymb,twocolumn]{revtex4}
\usepackage{amssymb}
\usepackage{amsmath}
\usepackage{amsfonts}
\usepackage{tikz}
\usepackage{bm}

\usepackage{epsfig}
\usepackage{t1enc}
\usepackage{soul}
\usepackage{color}

\usepackage{mathrsfs}
\usepackage{url}
\usepackage[all]{xy}
\usetikzlibrary{calc,patterns,angles,quotes}

\begin{document}

\title{Quantum principle of relativity}

\author{Andrzej Dragan}
\affiliation{Institute of Theoretical Physics, University of Warsaw, Pasteura 5, 02-093 Warsaw, Poland}
\affiliation{Centre for Quantum Technologies, National University of Singapore, 3 Science Drive 2, 117543 Singapore, Singapore}
\author{Artur Ekert}
\affiliation{Centre for Quantum Technologies, National University of Singapore, 3 Science Drive 2, 117543 Singapore, Singapore}
\affiliation{Mathematical Institute, University of Oxford, 24-29 St Giles’, OX1 3LB, UK}
\date{\today}

\begin{abstract}
We show that the local and deterministic mode of description is not only in conflict with the quantum theory, but also with relativity. We argue that elementary relativistic properties of spacetime lead to the emergence of a non-deterministic quantum-mechanical picture involving quantum superpositions and complex probability amplitudes. 
\end{abstract}

\maketitle

\section{Introduction}

Quantum theory is the most accurate description of reality that we currently possess. An agreement between theory's predictions and experimental data reaches an astounding precision of more than 10 digits \cite{Parker2018}. And yet several inventors of the theory, including Einstein, Schr\"{o}dinger, and de Broglie, doubted its correctness, because it painted a very disturbing image of reality. Even today, after nearly four decades of experiments violating Bell's inequalities \cite{Bell1964} many physicists are still puzzled by the quantum theory and often question our understanding of it.

There were many attempts to explore the quantum theory on a deeper level, for instance by deriving it from informational postulates \cite{Chiribella2011}. It is also known that correlations stronger than quantum might in principle exist and not be in conflict with relativity \cite{Popescu1994}. In order to recreate existing quantum correlations it is necessary to impose a requirement stronger than just "no-signaling", and specify, what is the maximum amount of information that can be extracted from a given volume of communication \cite{Pawlowski2009}. Other authors derive from first principles a quantum-mechanical rule for computing probabilities, known as the Born rule \cite{Zurek2005}, or even reinterpret the whole theory in a completely time-symmetric manner, using a peculiar approach known as the two-state formalism \cite{Vaidman2009} in which an amplitude of the process is dictated both by its past and its future.

But the most disturbing problem that Einstein and others were bothered with was not about what is the best set of axioms of the quantum theory or {\em how} to compute quantum probabilities. But rather {\em why} do we have to compute these probabilities in the first place, i. e. why reality on a microscopical scale is not deterministic and quantum particles, when not under direct observation, behave as if they existed at several locations at once. Unfortunately no theory so far succeeded in answering these questions on a more fundamental level. Even the string theory, which is sometimes considered to be a candidate for the ultimate theory, starts with an axiomatic string characterized by a quantum probability of splitting into two. Therefore even the string theory does not bring us any closer to understanding of the questions that bothered Einstein.

In this work we challenge this problem and suggest that the answer to these questions could have been in a plain sight for a while. We show how elementary special-relativistic considerations lead to the quantum paradigm, in which basic processes such as decays of elementary particles cannot happen at times that are locally predetermined \cite{Dragan2008}. In other words there can be no relativistic, local and deterministic theory predicting the moments of elementary quantum events, such as particle decays. We also show how relativistic considerations lead to the conclusion that the description of motion of a particle with just a single path is not possible and a quantum picture, in which multiple paths are involved is inevitable. 

An interplay between the quantum theory and relativity can lead to unintuitive new phenomena, such as indefinite causal structure \cite{Oreshkov2012} or superposition of spacetimes \cite{Bose2017}. We argue that similar phenomena appear in relativity alone, and the presence of relativistic structures, such as no-signaling, within the quantum theory is not a coincidence and becomes clear when a deeper connection between the quantum paradigm and relativity is revealed.

\section{All inertial observers\label{sec-relativity}}

We begin by deriving a generalized Lorentz transformation based on an illuminating observation of Ignatowsky, as well as Frank and Rothe, and Szymacha \cite{Ignatovsky1910,Pauli1981}, that the constancy of the speed of light as a postulate is not necessary to complete the derivation. 

Let us first consider a classical textbook $1+1$ dimensional case with two inertial frames $(t, x)$ and $(t', x')$ in a usual relative motion with the velocity $V$ of the primed frame with respect to the unprimed one. We are looking for the most general form of the transformation of coordinates between these two frames that is consistent with the Galilean principle of relativity. It follows that the only allowed transformations must be linear so that equations do not distinguish any instant of time or point in space. As a consequence all the transformation coefficients must be functions of the relative velocity $V$ only. It also follows that the inverse transformation involves a sign flip in the velocity $V$. Therefore we can write:
\begin{eqnarray}
\label{lineartransform}
x' &=& A(V) \,x +B(V)\,t,\nonumber \\
x  &=& A(-V)\,x'+B(-V)\,t',
\end{eqnarray}
where $A(V)$ and $B(V)$ are unknown functions we wish to determine. The origin of the primed frame of reference, given by the equation $x'=0$, is moving in the unprimed frame according to the equation $x=Vt$. Putting that into \eqref{lineartransform} we obtain the constraint: $\frac{B(V)}{A(V)} = -V$. This allows us to narrow down a family of possible transformations consistent with the principle of relativity to the following one:
\begin{eqnarray}
\label{lineartransform2}
x' & = & A(V)(x-Vt), \nonumber \\
t'  & = & A(V)\left(t-\frac{A(V)A(-V)-1}{V^2A(V)A(-V)}Vx\right),
\end{eqnarray}
where only a single function of velocity, $A(V)$ remains unknown. At this stage the only thing we can say about $A(V)$ is that it must be either a symmetric, or antisymmetric function of its argument. This is because a discrete change of sign of any spacetime coordinate in the unprimed frame should result in a discrete sign change in the transformation formulas \eqref{lineartransform2}. But since such a sign flip also affects the sign of velocity $V$, and consequently $A(V)$, therefore $A(V)$ can only be either symmetric, or anti-symmetric function of $V$.

In order to uniquely determine $A(V)$, let us consider a set of three inertial frames $(t, x)$, $(t', x')$, and $(t'', x'')$ in a relative motion. Let the primed frame move with the velocity $V_1$ relative to the unprimed frame, and let the double-primed frame move with the velocity $V_2$ relative to the primed one. By iterating the equations \eqref{lineartransform2} we obtain:
\begin{eqnarray}
\label{triple}
x'' &=&  A(V_1)A(V_2)x \left(1 + V_1 V_2
\frac{A(V_1)A(-V_1)-1}{V_1^2A(V_1)A(-V_1)} \right)\nonumber \\
& &- A(V_1)A(V_2)(V_1+V_2)t.
\end{eqnarray}
Looking at the structure of the first equation in \eqref{lineartransform2} we can compute the relative velocity $V$ by calculating the ratio between the coefficient in that transformation in front of $t$ and the coefficient in front of $x$ (and reversing the sign). Applying this rule to the formula \eqref{triple} we obtain:
\begin{eqnarray}
\label{veltrans}
V &=&  \frac{A(V_1)A(V_2)(V_1+V_2)}{1 + V_1 V_2
\frac{A(V_1)A(-V_1)-1}{V_1^2A(V_1)A(-V_1)} }.
\end{eqnarray}
Now, the crucial argument follows. If we interchange $V_1 \leftrightarrow -V_2$ in the above formula, we should obtain a velocity of the unprimed observer relative to the double-primed observer, which is just $-V$. Therefore:
\begin{eqnarray}
\label{veltrans2}
V &=&  \frac{A(-V_2)A(-V_1)(V_2+V_1)}{1 + V_2 V_1
\frac{A(-V_2)A(V_2)-1}{V_2^2A(-V_2)A(V_2)} }.
\end{eqnarray}
Whether $A(V)$ is symmetric or antisymmetric, we can drop negative signs in the arguments in the numerator of \eqref{veltrans2} and equate \eqref{veltrans2} with \eqref{veltrans}. Which brings us to the following condition:
\begin{equation}
\frac{A(V_1)A(-V_1)-1}{V_1^2A(V_1)A(-V_1)}=\frac{A(V_2)A(-V_2)-1}{V_2^2A(V_2)A(-V_2)}
\end{equation}
for any $V_1$ and $V_2$. This can only be satisfied if both sides of the equation are equal to some constant $K$:
\begin{equation}
\label{newconstant} \frac{A(V)A(-V)-1}{V^2A(V)A(-V)}=K,
\end{equation}
which sets another constraint on possible functions $A(V)$ appearing in \eqref{lineartransform2}. We are one step away from completing the derivation. 

For the symmetric case, $A(-V)=A(V)$, we can determine the form of $A(V)$ using \eqref{newconstant}, which leads to $A(V)=\pm\frac{1}{\sqrt{1-KV^2}}$. Choosing the sign such that for $V\to 0$ we get $x'\to x$, we retrieve familiar transformation formulas:
\begin{eqnarray}
\label{lineartransform3}
x' &=& \frac{x-Vt}{\sqrt{1-KV^2}}, \nonumber \\
t' &=& \frac{t-K V x}{\sqrt{1-KV^2}}.
\end{eqnarray}
The new constant $K$ characterizing fundamental properties of spacetime remains unknown. The case of $K=0$ corresponds to the Galilean universe, the case of $K>0$ leads to relativistic spacetime as we know it. The last case of $K<0$ corresponds to an Euclidean spacetime with one of the dimensions stretched by an extra factor of $\sqrt{|K|}$ and the derived transformation being just a regular rotation. From now on, we pick $K=\frac{1}{c^2}$, which brings us to the familiar formulas of the Lorentz transformation, that are well-behaved only for velocities $V<c$.

Let us now consider the anti-symmetric case of $A(-W)=-A(W)$, where we have chosen to denote the velocity with $W$ in order to discriminate it from the symmetric case. Using the constraint \eqref{newconstant} we retrieve the unique form of $A(W) = \pm\frac{W/|W|}{\sqrt{W^2/c^2-1}}$, which is well-behaved only for $W>c$ and leads to the following transformation formulas:
\begin{eqnarray}
\label{lineartransform4}
x' &=& \pm\frac{W}{|W|}\frac{x-Wt}{{\sqrt{W^2/c^2-1}}},\nonumber \\
t' &=& \pm\frac{W}{|W|}\frac{t-Wx/c^2}{{\sqrt{W^2/c^2-1}}}.
\end{eqnarray}
So far, we have only used the Galilean principle of relativity, which puts no restrictions on possible velocities of the observer. Both the solutions \eqref{lineartransform3} and \eqref{lineartransform4} preserve the constancy of the speed of light. In order to get rid of the second branch of solutions \eqref{lineartransform4}, we have to introduce additional physical assumptions that rule them out. We will choose not to do so, and instead we will investigate what are the consequences of the existence of these extra solutions. The purpose of this work is to show that keeping the second branch of solutions leads to the change of principles of causality, however not as logical-inconsistencies, but rather in the form of a non-deterministic behavior known from the quantum theory.

A few comments are in order. First, let us note that the both sets of equations \eqref{lineartransform3} and \eqref{lineartransform4} preserve the speed of light, so any derivation of the Lorentz transformations should also lead to the possible second branch of solutions given by \eqref{lineartransform4}. If not then either something is overlooked, or additional limiting assumptions are taken. Second, the sign in front of the equations \eqref{lineartransform4} cannot be uniquely determined, because no $W\to 0$ limit exists. The choice of the sign must remain a matter of convention, and from now on we will pick the negative sign. This however does not imply that the extra antisymmetric term $\frac{W}{|W|}$ can be skipped. It turns out that without that term the theory looses its relativistic invariance, although some authors make a mistake of forgetting it. The first appearance of the correct formula \eqref{lineartransform4} in the literature can be found in \cite{Machildon1983}. Third, both branches of solutions form a group structure only in the considered $1+1$ dimensional scenario. This is not the case in the $1+3$ dimensional case \cite{Machildon1983a}, therefore we will carefully discuss this case separately in the further part of this paper. For now we stick to the $1+1$ scenario and investigate its consequences.

\begin{figure}
\begin{center}
\epsfig{file=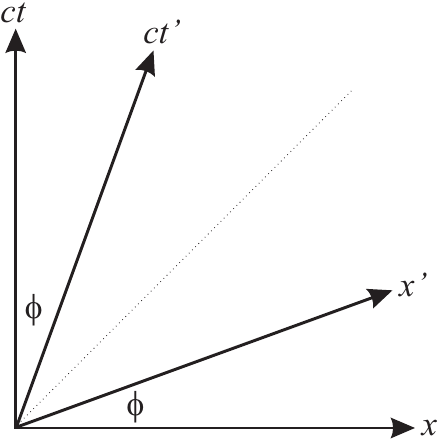} \caption{ \label{hiperrot} \sf
\footnotesize Hyperbolic rotation by an angle $\phi$ corresponding to a subluminal transformation. }
\end{center}
\end{figure}

Finally, let us also comment on the geometrical interpretation of the derived transformation formulas \eqref{lineartransform3} and \eqref{lineartransform4}. The standard result involving the subluminal branch of solutions \eqref{lineartransform3} corresponds to a hyperbolic rotation by an angle $-\frac{\pi}{4}<\phi<\frac{\pi}{4}$, as shown in Fig.~\ref{hiperrot}. The second branch of solutions \eqref{lineartransform4} is also a hyperbolic rotation, but by the angle $\frac{\pi}{4}<\phi<\frac{3\pi}{4}$. Note that thanks to the antisymmetric term $\frac{W}{|W|}$ appearing in \eqref{lineartransform4}, the superluminal branch of transformations forms an ortochronous structure with a well-defined direction of time.

\section{Indeterministic behavior}

So far we have shown that the Galilean principle of relativity alone leads to two branches of coordinate transformations corresponding to subluminal and superluminal families of observers. In the $1+1$ dimensional scenario considered so far, these branches are indistinguishable, which means that a particle at rest with respect to an observer belonging to one of the branches will be considered superluminal by the observer belonging to the other branch. In other words, being superluminal is a relative property. Let us investigate the new aspects stemming from the fact that we take both branches of solutions into consideration. We will first show that relativistic, local, and deterministic mode of description of fundamental processes is no longer possible.

Suppose that a superluminal particle observed by some inertial observer has been emitted from a source particle at the event {\sf A} and then absorbed at some later time by the identical target particle at the event {\sf B} - as shown in Fig.~\ref{Procesy2}a \footnote{In this and the following figures we pick a convention of depicting subluminal particles with solid lines, and superluminal ones with dashed lines.}. Energy-momentum conservation allows for such a process to occur, as we  show later. The same process observed from a reference frame moving with a relative subluminal velocity is depicted in Fig.~\ref{Procesy2}b. In this frame the event {\sf B} becomes the emission of the superluminal particle, while {\sf A} becomes the absorption. 
\begin{figure}
\begin{center}
\epsfig{file=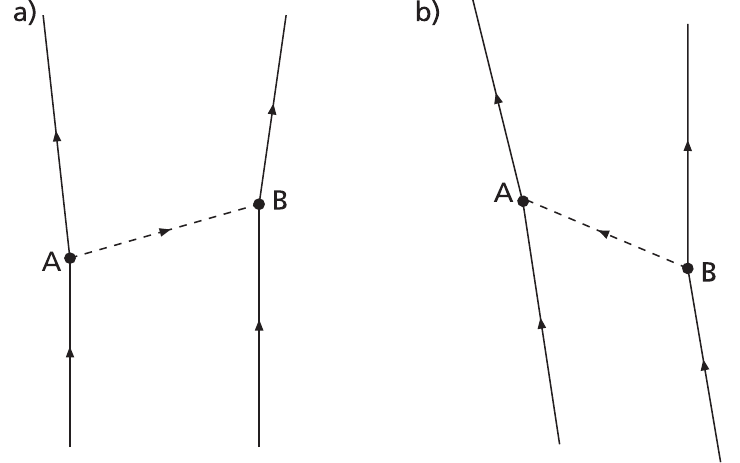} \caption{ \label{Procesy2} \sf
\footnotesize Spacetime diagrams of a process of sending a superluminal particle
as seen by two inertial observers (time is vertical, space is horizontal): a) particle emitted from A and
absorbed in B, b) the same process observed in a different inertial frame.}
\end{center}
\end{figure}

Let us focus on the first frame shown in Fig.~\ref{Procesy2}a and assume that the moment of emission at {\sf A} could be predicted using a local and deterministic mode of description. In other words, let us assume that the past world-line of the source particle prior to the event {\sf A} contains locally all the information necessary to predict the exact moment of emission of a superluminal particle at {\sf A}. Or using the Einsteinian language, there is an element reality to it. On the other hand someone holding the target particle {\sf B} cannot predict the moment of the absorption at {\sf B} based only on local measurements of the particle {\sf B} prior to the event. Now, let us change the reference frame and study the same scenario from the perspective of the observer depicted in Fig.~\ref{Procesy2}b. Let us try to answer the following question: what caused the emission of the superluminal particle at the event {\sf B}?

We could answer by saying that the cause of the event {\sf B} takes place in the distant world line of the particle {\sf A}. Possibly at a later time than the event {\sf B} itself. However, if we seek a deterministic {\em and} local mode of description, i.~e. try to determine the moment of emission at {\sf B} only by a local measurement on the particle {\sf B}, it is clearly impossible. We have already assumed that the past world-line of the particle {\sf B} carries no information about the time of the event {\sf B}. In practice, the observer having only access to the local properties of the particle {\sf B} can only conclude that the emission at {\sf B} was be completely spontaneous and fundamentally unpredictable.

We have previously assumed that the cause of the emission of the superluminal particle at {\sf A} (in the first reference frame) was determined by the past world-line of {\sf A}. This assumption leads, however to a preferred reference frame, in which a local deterministic mode of description is possible, while it remains impossible in other frames. In order to preserve the Galilean principle of relativity stating that no preferred inertial reference frame exists, we have to abandon our assumption that the emission at {\sf A} in the first frame could be determined by a local process. As a result we conclude that no relativistic, local and deterministic description of the emission of a superluminal particle is possible in any inertial frame. If such an emission was to take place, it would have to appear completely random to any inertial observer. If we had a source of superluminal particles at our disposal, we would not be able to use it to send any information because we would not be able to control the emission rate using any local operations.

\begin{figure}
\begin{center}
\epsfig{file=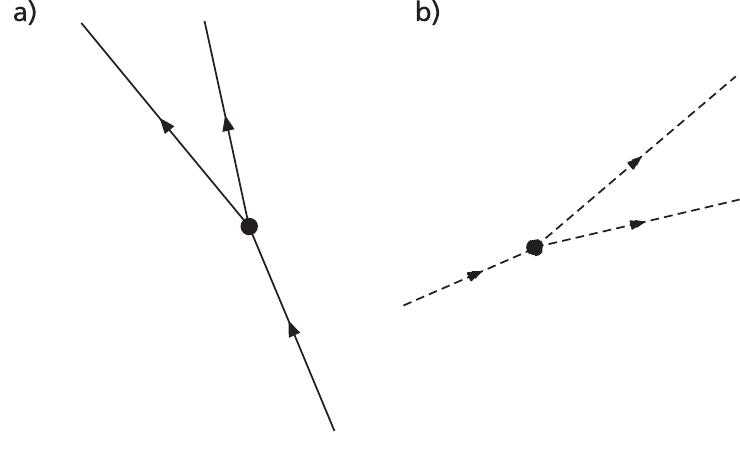} \caption{ \label{Procesy3} \sf
\footnotesize A spacetime diagram of a decay of a subluminal particle into a pair of subluminal particles (time is vertical, space is horizontal): a) in a subluminal reference frame, b) in a superluminal reference frame.}
\end{center}
\end{figure}

Non-deterministic behavior is not only a property of superluminal particles. The same applies to subluminal particles, which can be shown in the following way. Consider a decay of a subluminal particle into a pair of other subluminal particles, as depicted in Fig.\ref{Procesy3}a. Let us picture the same process as seen by the infinitely fast moving inertial observer, for which the transformation \eqref{lineartransform4} reduces to:
\begin{eqnarray}
\label{inftransform}
x' &=& ct,\nonumber \\
ct' &=& x.
\end{eqnarray}
For such a frame, the considered decay process is depicted in Fig.~\ref{Procesy3}b. In this frame all particles involved in the process are superluminal and henceforth, the decay cannot be described using any local and deterministic theory, as we have shown earlier. By invoking the Galilean principle of relativity we conclude that the same must be the case for any subluminal reference frame.

\section{Multiple paths}
Another fundamentally axiomatic property of the quantum theory, besides it being non-deterministic, is the fact that a particle that is not being observed can behave as if it was moving along multiple trajectories at once, which is best shown in interference experiments. But once the particle is observed it can only be detected at one of the locations. Now, let us show how this follows from the Galilean principle of relativity involving both families of inertial observers \eqref{lineartransform3} and \eqref{lineartransform4}.

\begin{figure}
\begin{center}
\epsfig{file=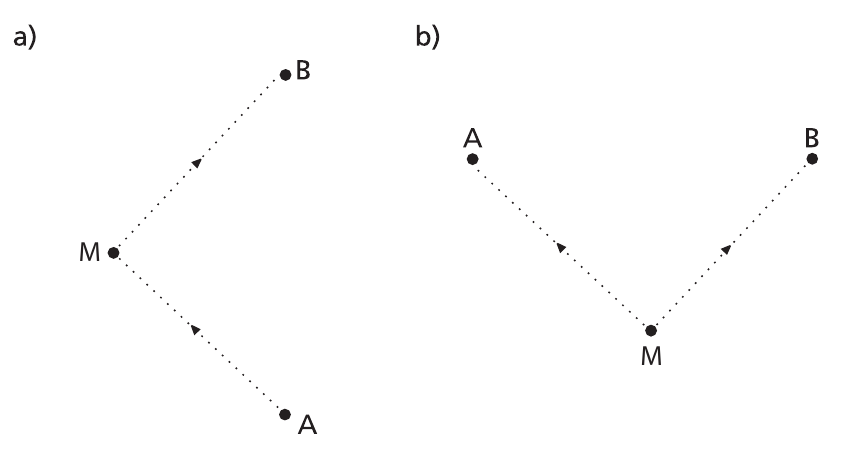} \caption{ \label{Procesy4} \sf
\footnotesize A spacetime diagram of a luminal particle (dotted line) reflected from a mirror (time is vertical, space is horizontal): a) in a subluminal reference frame, b) in a superluminal reference frame.}
\end{center}
\end{figure}

Consider a photon emitted from a source at {\sf A}, reflected from a mirror {\sf M} and then received at {\sf B}, as shown in Fig.~\ref{Procesy4}a. Suppose that we want to detect the photon by placing detectors at its path. If a detector placed at the path {\sf A-M} detects the photon and absorbs it, then a similar detector placed at the path {\sf M-B} will not register anything, because the photon  has been absorbed earlier. Similarly, if a detector at {\sf M-B} absorbed the photon, then certainly, the photon could not have been detected at the path {\sf A-M}. Now let us analyze the same scenario from an infinitely fast moving reference frame by applying the transformation equations \eqref{inftransform}. In this reference frame the photon is traveling from {\sf M} towards {\sf A} and {\sf B} along two paths, but if we try to detect it using a pair of detectors placed at {\sf M-A} and {\sf M-B} then only one of them can absorb the photon. However as long as we do not make any observation, the motion of the photon is characterized by two parallel paths, not one.

As we can see, even if we start with an idea of a classical particle moving along a single path, it is only a matter of a change of the reference frame to arrive at a scenario involving more than one path. 

\begin{figure}
\begin{center}
\epsfig{file=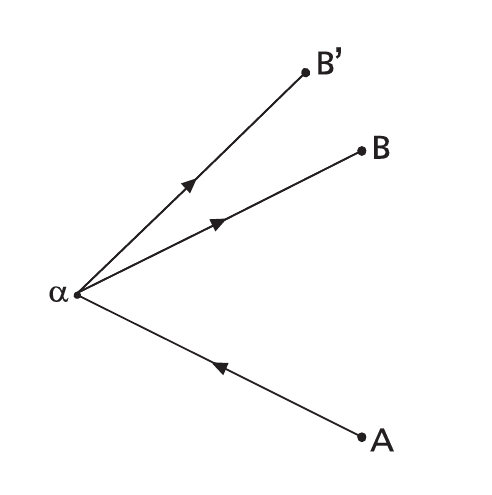} \caption{ \label{Procesy5} \sf A particle emitted in A is scattered at $\alpha$ into a motion along two paths towards B and B' at once.
\footnotesize }
\end{center}
\end{figure}

Consider a process depicted in Fig.~\ref{Procesy5} in which a particle emitted in {\sf A} is scattered in $\alpha$, where it starts to follow two paths at once towards {\sf B} and {\sf B'}. The same process viewed from the infinitely fast moving frame will involve the particle following three paths at once. This concept can be iterated leading to scenarios involving multiple paths at once. Once both branches of transformations \eqref{lineartransform3} and \eqref{lineartransform4} are involved, a classical description of a particle always moving along a single trajectory becomes inconsistent with the Galilean principle of relativity.

\section{Probability amplitudes}
\begin{figure}
\begin{center}
\epsfig{file=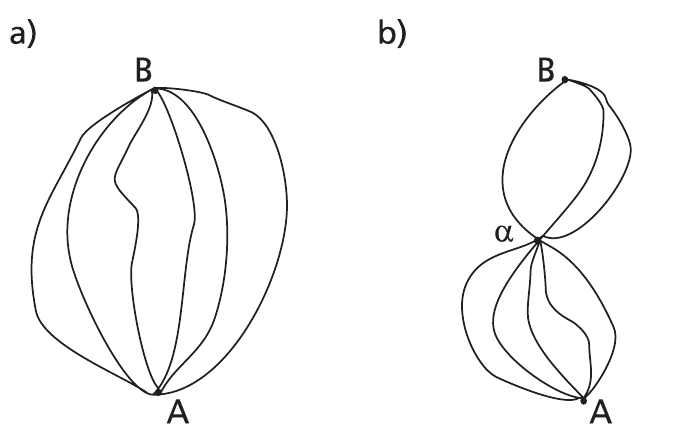} \caption{ \label{Procesy6} \sf Multiple paths connecting two spacetime points, A and B, in two possible settings.
\footnotesize }
\end{center}
\end{figure}

Let us go back to the orthodox question of subluminal particles being observed by subluminal observers. Relativistic theories operate on notions that do not change under Lorentz transformations. We will therefore investigate, what are relativistically invariant quantities that describe a particle motion between two points, {\sf A} and {\sf B}. The simplest relativistic invariant characterizing a single path is its relativistic length, i.e. the proper time:
\begin{equation}
\label{phase} 
\phi \sim \int_{\sf A}^{\sf B} \sqrt{1-v^2/c^2}\text{d}t
\sim\int_{\sf A}^{\sf B}(E\,\mbox{d}t -
p\, \mbox{d}x),
\end{equation}
where $E\sim \frac{1}{\sqrt{1-v^2/c^2}}$ is the energy of the particle and $p\sim\frac{v}{\sqrt{1-v^2/c^2}}$ is its momentum. We will choose the proportionality constant such that the relativistic invariant $\phi$ is dimensionless and will be referred to as the phase along the path. When multiple paths are involved, as shown in Fig.~\ref{Procesy6}a, a relativistic invariant characterizing such a diagram must be a function of phases along all individual paths, ${\cal P}^{(n)} (\phi_1, \phi_2, \ldots,\phi_n)$, where ${\cal P}$ is a smooth function and $n$ is the number of possible paths. We will be interested in the question, what are reasonable functions ${\cal P}$ to consider. It turns out that the family of such functions is not too vast.

We will spell out three basic requirements that we impose on possible functions ${\cal P}$ and then study, what are we left with. First of all, our choice of the way we label individual paths has no physical significance, therefore a reasonable function ${\cal P}$ should be a symmetric function of its arguments:
\begin{equation}
\label{axiom-symmetry} {\cal P}^{(n)} (\phi_1, \phi_2,
\ldots,\phi_n) = {\cal P}^{(n)} (\phi_{\pi(1)}, \phi_{\pi(2)},
\ldots,\phi_{\pi(n)}),
\end{equation}
where $\pi$ is an any permutation of an $n$-element set. Second of all, we choose our description to be completely time-symmetric, which means that flipping the sign of all the phases should not affect ${\cal P}$:
\begin{equation}
\label{axiom-inverse} {\cal P}^{(n)} (\phi_1, \phi_2,
\ldots,\phi_n) = {\cal P}^{(n)} (-\phi_1, -\phi_2,
\ldots,-\phi_n).
\end{equation}
Consider a special type of motion, in which all trajectories intersect at a single point $\alpha$, as shown in Fig.\ref{Procesy6}b. Suppose that the events {\sf A} and $\alpha$ are linked by $n$ different paths characterized by phases $\phi_1, \ldots, \phi_n$, while $\alpha$ and {\sf B} are interlinked by $m$ paths  characterized by phases $\xi_1, \ldots, \xi_m$. The total number of paths connecting {\sf A} and {\sf B} is equal to $nm$ and since the phases are additive, such motion involves sums of phases $\phi_i + \xi_j$. Therefore the invariant function for such a motion is ${\cal P}^{(nm)}(\phi_1+\xi_1, \phi_1+\xi_2, \phi_1+\xi_3, \ldots, \phi_n+\xi_m)$. 

Our third and final condition captures the fact that the invariant quantity we are trying to establish should have properties of probability. Since the probability for the particle to travel from {\sf A} to {\sf B} should be a product of probabilities for the particle to travel from {\sf A} to $\alpha$ and then from $\alpha$ to {\sf B}, we impose the third condition:
\begin{widetext}
\begin{equation}
\label{axiom-probability}  {\cal P}^{(nm)}(\phi_1+\xi_1, \phi_1+\xi_2, \phi_1+\xi_3, \ldots, \phi_n+\xi_m) = {\cal P}^{(n)} (\phi_1, \phi_2, \ldots,\phi_n){\cal P}^{(m)} (\xi_1, \xi_2, \ldots,\xi_m).
\end{equation}
One special case that satisfies all the conditions \eqref{axiom-symmetry}, \eqref{axiom-inverse} and \eqref{axiom-probability} has the form:
\begin{equation}
\label{AEfunction} {\cal P}^{(n)}(\phi_1, \phi_2, \ldots, \phi_n)
= \frac{1}{n^\beta} \left( e^{\alpha\phi_1} + e^{\alpha\phi_2} + \ldots + e^{\alpha\phi_n}\right)^\gamma 
\left( e^{-\alpha\phi_1} + e^{-\alpha\phi_2} + \ldots + e^{-\alpha\phi_n}\right)^\gamma,
\end{equation}
\end{widetext}
where $\alpha$, $\beta$ and $\gamma$ are arbitrary constants. In the Appendix~\ref{proof-wavefunction} we show that a general form of the function ${\cal P}$ is a multi product of special solutions \eqref{AEfunction} with any constants $\alpha$, $\beta$ and $\gamma$. 

For the infinite number of trajectories all these invariants tend to diverge if $\gamma>0$ or go to zero if $\gamma<0$. The invariant can remain finite for the infinite number of paths only if the constant $\alpha$ takes a purely imaginary value. As a consequence, a relativistically invariant description of the scenario, in which a particle is moving along (infinitely) many possible paths involves the following quantity:
\begin{equation}
\langle {\sf B} | {\sf A}\rangle \sim \sum_k e^{i|\alpha|\phi_k},
\end{equation}
known as the (complex) probability amplitude, for which the proportionality constant can be established based on the normalization condition.

This result shows that relativistic invariance and symmetry requirements lead to the characterization of the probability-like quantities that are based on a sum of complex exponential functions that we call probability amplitudes.

\section{$1+3$ dimensional case}
Galilean principle of relativity for the $1+1$ dimensional spacetime involving two families of inertial observers leads to processes such as particle decays, that violate a classical, deterministic and local mode of description. Moreover, particle motions inevitably involve multiple trajectories at once, which combined with the relativistically invariant description can be best summarized using the quantum principle of superposition.

The situation becomes even more interesting for the $1+3$ dimensional case. It has been shown that the smallest group involving both subluminal and superluminal four-dimensional transformations is $SL(4,\mathbb{R})$ \cite{Machildon1983a}. It is clear that this cannot be a symmetry group, because it involves transformations such as direction-dependent time dilation, which are not observed \cite{Machildon1983a}. This suggests that superluminal transformations in the $1+3$ dimensional spacetime should not be symmetries at all. A possible interpretation of this result has been given in \cite{Sutherland1986}, where the authors suggest that unlike the $1+1$ dimensional case, the family of superluminal observers can be distinguished from the subluminal observers and therefore being superluminal is not a relative notion anymore. A physical justification given by the authors is the following. Let us extend the coordinate transformation \eqref{lineartransform4} by adding the trivial perpendicular counterparts, $y'=y$ and $z'=z$. The spacetime interval in the new coordinates now becomes:
\begin{equation}
\label{interval}
c^2\text{d}t^2-\text{d}x^2-\text{d}y^2-\text{d}z^2=-c^2\text{d}t'^2+\text{d}x'^2-\text{d}y'^2-\text{d}z'^2.
\end{equation}
The authors of \cite{Sutherland1986} conclude that the spatial component $\text{d}x'^2-\text{d}y'^2-\text{d}z'^2$ defines a non-Euclidean space, which can be physically discriminated from the Euclidean space of the subluminal observers and therefore there exist physical differences between subluminal and superluminal inertial observers. 

Since the Galilean principle of relativity stating that all inertial frames are equivalent  does not hold in the $1+3$ dimensional spacetime, we propose a quantum version of the principle of relativity. Let us postulate that the existence or non-existence of a local and deterministic mode of description of any process should not depend on the choice of the inertial reference frame. For example if there is no local deterministic mechanism (or "element of reality") behind the particle decay in Fig.~\ref{Procesy2}b in the past world-line of {\sf B} in one frame, there should be no such mechanism in any other frame. In this way all the conclusions of the previous sections are still valid, while we allow for the two families of observers to be physically distinguishable.

Lastly, we would like to propose a different interpretation of the relation between spacetime intervals in subluminal and superluminal reference frames, given by \eqref{interval}. Let us notice, that the common signs of individual terms on the right-hand side of the equation \eqref{interval} suggest that the temporal coordinate $\text{d}t'$ should have the same properties as $\text{d}y'$ and $\text{d}z'$. The quantity $\text{d}t'$ can be identified as a temporal coordinate, because its axis $t'$ must coincide with the world line of the superluminal observer. This suggests, that the remaining coordinates, $y'$ and $z'$ are also temporal, and there is only a single spatial dimension in a superluminal frame of reference, $x'$. Within such an interpretation, the interval in the $n+m$ dimensional spacetime, defined as:
\begin{equation}
\text{d}s^2 \equiv c^2 \sum_{i=1}^n \text{d} t_i^2 - \sum_{i=1}^m \text{d} r_i^2
\end{equation}
changes its sign for the superluminal coordinate transformation, and the two perpendicular spatial coordinates change their character transforming the $1+3$ dimensional spacetime into the $3+1$ dimensional one.

Such a peculiar property of superluminal observers, not only explains a physical difference between them and subluminal reference frames, but also offers an interesting insight into the origins of the wave properties of matter.

In a scenario involving a more than one temporal dimension one has to expect that objects would not move along single world lines (which would violate rotational symmetry), but instead they would age in three dimensions, propagating in all of them. But such peculiar dynamics observed from a subluminal frame of reference would look just as a propagation along all possible spatial dimensions. Such a behavior takes place in reality and it is sometimes described using the Huygens principle stating that {\em any} point, in which a physical particle is placed is a source of a new spherical ,,mater wave''. Indeed, one of the consequences of our interpretation of the expression \eqref{interval} would be that all physical objects should undergo the Huygens principle and propagate as spherical waves from every point at which they arrive.

The $1+3$ dimensional Lorentz transformation between two subluminal observers is obtained from \eqref{lineartransform3} by replacing $Vx$ with $\boldsymbol{V}\cdot\boldsymbol{r}$, where $\boldsymbol{V}$ is an arbitrary subluminal velocity and $\boldsymbol{r} = (x, y, z)$. It can be written in the coordinate-independent form as:
\begin{eqnarray}
\boldsymbol{r}' &=& \boldsymbol{r}-\frac{\boldsymbol{V}\cdot\boldsymbol{r}}{V^2}\boldsymbol{V} + \frac{\frac{\boldsymbol{V}\cdot\boldsymbol{r}}{V^2}-t}{\sqrt{1-V^2/c^2}}\boldsymbol{V},\nonumber \\
ct'  &=& \frac{ct-\frac{\boldsymbol{V}\cdot\boldsymbol{r}}{c}}{\sqrt{1-V^2/c^2}}.
\end{eqnarray}
The inverse transformation is obtained by substituting $\boldsymbol{V}\to-\boldsymbol{V}$, as well as $\boldsymbol{r}\leftrightarrow\boldsymbol{r'}$ and $t\leftrightarrow t'$. 

Similar generalization can be carried out for the superluminal transformations \eqref{lineartransform4}. By replacing $Wx$ in \eqref{lineartransform4} with $\boldsymbol{W}\cdot\boldsymbol{r}$, we obtain the coordinate-independent transformation between a subluminal reference frame $(t,\boldsymbol{r})$ and a superluminal one $(\boldsymbol{t'},x')$ moving with a superluminal velocity $\boldsymbol{W}$:

\begin{eqnarray}
x'  &=& \frac{Wt-\frac{\boldsymbol{W}\cdot\boldsymbol{r}}{W}}{\sqrt{W^2/c^2-1}},\nonumber \\
c\boldsymbol{t}' &=& \boldsymbol{r}-\frac{\boldsymbol{W}\cdot\boldsymbol{r}}{W^2}\boldsymbol{W} + \frac{\frac{\boldsymbol{W}\cdot\boldsymbol{r}}{Wc}-\frac{ct}{W}}{\sqrt{W^2/c^2-1}}\boldsymbol{W}.
\end{eqnarray}
The inverse transformation is obtained by reversing the above set of linear equations. It is equivalent to substituting $\boldsymbol{W}\to-\boldsymbol{W}$, as well as $\boldsymbol{r}\leftrightarrow\boldsymbol{t'}$ and $t\leftrightarrow x'$. For the infinite speed limit $W\to\infty$, the above formulas reduce to:
\begin{eqnarray}
x'  &=&ct,\nonumber \\
c\boldsymbol{t}' &=& \boldsymbol{r},
\end{eqnarray}
regardless of the direction of the velocity $\boldsymbol{W}$.

\section{Summary}
Ruling out from special relativity unwanted superluminal family of observers is not necessary, but results in a classical description of a particle moving along a well-defined single trajectory that is in conflict with the predictions of the quantum theory (and experiments). Keeping both families of solutions instead leads to a scenario involving non-deterministic behavior and non-classical motion of particles as a straightforward and natural consequence. Such an approach reveals a connection between special relativity and quantum theory that reaches much deeper than previously thought. It involves several unintuitive consequences and challenges our understanding of basic concepts of space and time, but offers a clear justification of some of the most intriguing axioms of the quantum theory, including non-deterministic behavior and wave-like dynamics of matter.

\begin{acknowledgements} 
We thank Iwo Bia{\l}ynicki-Birula, Ryszard Horodecki, Bogdan Mielnik, Holger Nielsen, and Sandu Popescu for insightful conversations. 
\end{acknowledgements}

\begin{appendix}

\section{Derivation of all the probability-like relativistic invariants\label{proof-wavefunction}}

Let ${\cal P}^{(n)}(\phi_1, \phi_2, \ldots, \phi_n)$ and ${\cal R}^{(n)}(\phi_1, \phi_2, \ldots, \phi_n)$ be arbitrary smooth functions obeying all the conditions \eqref{axiom-symmetry}, \eqref{axiom-inverse}, and \eqref{axiom-probability}. We find that the product of arbitrary powers of ${\cal P}^{(n)}(\phi_1, \phi_2, \ldots, \phi_n)$ and ${\cal R}^{(n)}(\phi_1, \phi_2, \ldots, \phi_n)$ is also smooth and obeys the above axioms. Therefore a product, ratio or any power of any special solutions to the problem is also a solution. Similarly, one can verify that a sum of non-trivial solutions ${\cal P}^{(n)}(\phi_1, \phi_2, \ldots, \phi_n) + {\cal R}^{(n)}(\phi_1, \phi_2, \ldots, \phi_n)$ does not satisfy the condition \eqref{axiom-probability} therefore it is not a valid solution.

Let us consider a Taylor expansion of the smooth, completely symmetric function ${\cal P}^{(n)}(\phi_1, \phi_2, \ldots, \phi_n)$. A collection of terms in each order of expansion is a completely symmetric polynomial. According to Cauchy's theorem \cite{Sierpinski1946}, such polynomials can be decomposed into symmetric polynomials, defined as $E^{(k)}(\phi_1, \phi_2, \ldots, \phi_n)\equiv\sum_{i=1}^n \phi_i^k$, so that the Taylor expansion takes the following form:
\begin{equation}
\label{expansion}
\begin{split}
{\cal P}^{(n)}(\phi_1, \phi_2, \ldots, \phi_n) = \sum_{l=0}^\infty
\sum_{k_1,k_2,\ldots,k_{l}=1}^\infty
\alpha^{(n)}_{k_1,k_2,\ldots,k_l} \\
\times E^{(k_1)}(\phi_1,\phi_2,\ldots,\phi_n)\cdots
E^{(k_l)}(\phi_1,\phi_2,\ldots,\phi_n),
\end{split}
\end{equation}
with some collection of coefficients $\alpha^{(n)}_{k_1,k_2,\ldots,k_l}$. The symmetric polynomials $E^{(k)}(\phi_1, \phi_2, \ldots, \phi_n)$ are algebraically independent for $k<n$ \cite{Sierpinski1946}, but we will be interested in the large $n$ limit, in which case they all can be treated as algebraically independent.

Let us first consider a special case of the expansion \eqref{expansion}, in which only the terms with a fixed value of $l=N$ do not vanish. In this case we can skip the summation over $l$ in the Taylor expansion \eqref{expansion}, which reduces the expression to:
\begin{equation}
\label{expansion4}
\begin{split}
{\cal P}^{(n)}(\phi_1, \phi_2, \ldots, \phi_n) = \sum_{k_1, k_2,
\ldots,k_N=0}^\infty \alpha^{(n)}_{k_1, k_2, \ldots,k_N} \\
E^{(k_1)}(\phi_1,\phi_2,\ldots,\phi_n)\cdots
E^{(k_N)}(\phi_1,\phi_2,\ldots,\phi_n).
\end{split}
\end{equation}
Inserting this into the condition \eqref{axiom-probability} yields:
\begin{widetext}
\begin{equation}
\begin{split}
\label{singleE} &\sum_{r_1, r_2,
\ldots,r_N=0}^\infty \alpha^{(nm)}_{r_1, r_2, \ldots,r_N}
E^{(r_1)}(\phi_1+\xi_1,\phi_1+\xi_2,\ldots,\phi_n+\xi_m)\cdots E^{(r_N)}(\phi_1+\xi_1,\phi_1+\xi_2,\ldots,\phi_n+\xi_m) \\
& =
\sum_{k_1, k_2,
\ldots,k_N=0}^\infty \alpha^{(n)}_{k_1, k_2, \ldots,k_N}
E^{(k_1)}(\phi_1,\phi_2,\ldots,\phi_n)\cdots E^{(k_N)}(\phi_1,\phi_2,\ldots,\phi_n) \\
& \times \sum_{s_1, s_2,\ldots,s_N=0}^\infty \alpha^{(m)}_{s_1, s_2, \ldots,s_N} E^{(s_1)}(\xi_1,\xi_2,\ldots,\xi_m)\cdots E^{(s_N)}(\xi_1,\xi_2,\ldots,\xi_m).
\end{split}
\end{equation}
Terms appearing in the left hand side of \eqref{singleE} can be expanded using the definition of $E^{(n)}$ and the Newton's power formula:
\begin{equation}
\label{newtexp}
\begin{split}
E^{(r_i)}(\phi_1+\xi_1,\phi_1+\xi_2,\ldots,\phi_n+\xi_m) = \sum_{t_i=0}^{r_i}\binom{r_i}{t_i}E^{(t_i)}(\phi_1,\phi_2,\ldots,\phi_n)
E^{(r_i-t_i)}(\xi_1,\xi_2,\ldots,\xi_m).
\end{split}
\end{equation}
Inserting \eqref{newtexp} into \eqref{singleE} and invoking the algebraic independence of the polynomials $E^{(k)}$ we obtain the condition for the coefficients $\alpha^{(n)}_k$:
\begin{equation}
\sum_{\pi,\pi'} \binom{k_{\pi(1)}+s_{\pi'(1)}}{k_{\pi(1)}}\cdots
\binom{k_{\pi(N)}+s_{\pi'(N)}}{k_{\pi(N)}}\,
\alpha^{(nm)}_{k_{\pi(1)}+s_{\pi'(1)},\ldots,k_{\pi(N)}+s_{\pi'(N)}} = 
\sum_{\sigma,\sigma'}\alpha^{(n)}_{k_{\sigma(1)},\ldots,k_{\sigma(N)}}
\alpha^{(m)}_{s_{\sigma'(1)},\ldots,s_{\sigma'(N)}},
\end{equation}
where $\sigma$, $\sigma'$, $\pi$, and $\pi'$ are arbitrary
permutations of an $N$-element set. Without a loss of generality
we can assume that the coefficients $\alpha^{(n)}_{k_1, k_2,
\ldots,k_N}$ are completely symmetric in their indices $k_i$, because
any nonsymmetric component does not contribute to the overall sum
\eqref{expansion4} anyway. This assumption yields:
\begin{equation}
\label{Cauchy} N! k_1! k_2!\cdots k_N! s_1! s_2!\cdots s_N!\,
\alpha^{(n)}_{k_1,\ldots,k_N} \alpha^{(m)}_{s_1,\ldots,s_N} =
\sum_{\pi} (k_1+s_{\pi(1)})!\cdots (k_N+s_{\pi(N)})!\,
\alpha^{(nm)}_{k_1+s_{\pi(1)},\ldots,k_N+s_{\pi(N)}},
\end{equation}
\end{widetext}
with the following solution:
\begin{equation}
\label{alpha2} \alpha^{(n)}_{k_1,k_2,\ldots,k_N} =
\frac{1}{n^{\beta'}}
\frac{\sum_\pi\alpha_{1}^{k_{\pi(1)}}\alpha_{2}^{k_{\pi(2)}}\cdots
\alpha_{N}^{k_{\pi(N)}}}{N!k_1! k_2!\cdots k_N!},
\end{equation}
where $\alpha_1, \alpha_2, \ldots, \alpha_N$ are arbitrary
constants. This special solution satisfying the 
axioms \eqref{axiom-symmetry} and \eqref{axiom-probability} can be explicitly written by
substituting \eqref{alpha2} into \eqref{expansion4}:
\begin{widetext}
\begin{equation}
\begin{split}
{\cal P}^{(n)}(\phi_1, \phi_2, \ldots, \phi_n) & =
\frac{1}{n^{\beta'}} \sum_{k_1, k_2,
\ldots,k_N=0}^\infty
\frac{\sum_\pi\alpha_{1}^{k_{\pi(1)}}\alpha_{2}^{k_{\pi(2)}}\cdots
\alpha_{N}^{k_{\pi(N)}}}{N!k_1! k_2!\cdots k_N!}
E^{(k_1)}(\phi_1,\phi_2,\ldots,\phi_n)\cdots
E^{(k_N)}(\phi_1,\phi_2,\ldots,\phi_n) \\
& = \frac{1}{n^{\beta'}} \sum_{k_1, k_2,
\ldots,k_N=0}^\infty \frac{\alpha_1^{k_1}\alpha_2^{k_2}\cdots
\alpha_N^{k_N}}{k_1! k_2!\cdots k_N!}
E^{(k_1)}(\phi_1,\phi_2,\ldots,\phi_n)\cdots
E^{(k_N)}(\phi_1,\phi_2,\ldots,\phi_n) \\
& = \frac{1}{n^{\beta'}}
\left(e^{\alpha_1\phi_1}+e^{\alpha_1\phi_2}+\ldots+e^{\alpha_1\phi_n}\right)\cdots
\left(e^{\alpha_N\phi_1}+e^{\alpha_N\phi_2}+\ldots+e^{\alpha_N\phi_n}\right).
\end{split}
\end{equation}
\end{widetext}
Taking into account the condition \eqref{axiom-inverse} we conclude that for each parameter $\alpha_i$ the last product must contain another term with a negative parameter $-\alpha_i$, therefore $N$ must be an even number. We also recall that the sum of solutions to our problem is not a solution, therefore the summation over $l$ in the formula \eqref{expansion} can be skipped. The special solution \eqref{AEfunction} corresponds to the simplest case of $N=2$.

\end{appendix}

\end{document}